\documentclass[floatfix,11pt,onecolumn,showpacs,amsmath,amssymb]{revtex4}

\usepackage{epsf}
\usepackage{graphicx}  
\usepackage{dcolumn}   
\usepackage{bm}        
\usepackage{color}

\newcommand{\be}{\begin{equation}}
\newcommand{\en}{\end{equation}}
 \newcommand{\bea}{\begin{eqnarray}}
 \newcommand{\ena}{\end{eqnarray}}

\begin{document}

\title{Non perturbative spherical gravitational waves}
\author{ Hongsheng Zhang$^{1}$\footnote{Electronic address: sps\_zhanghs@ujn.edu.cn} }
\affiliation{$^1$ School of Physics and Technology, University of Jinan, 336 West Road of Nan Xinzhuang, Jinan, Shandong 250022, China}

\date{ \today}

\begin{abstract}
   It is well-known that there is no spherical/topologically spherical gravitational waves  in vacuum space in general relativity. We show that a deviation from general relativity leads to exact vacuum spherical gravitational waves, no matter how tiny this deviation is. We also discuss the related topics, including Vaidya-like metric in $f(R)$ gravity. We demonstrate that the existence of spherical gravitational wave is a non perturbative property for gravities. We investigate energy carried by this nonperturbative wave. We first find the wave solution from investigations of Vaidya-like metric in $f(R)$ gravity, which has only one longitude polarization. We further extend it to a metric with two transverse polarizations by directly solving the field equation.

\end{abstract}


\preprint{arXiv: }
 \maketitle

  A general property of modern field theory is that all interactions propagate at finite velocity in manner of waves. Almost immediately after the construction of general relativity, Einstein proposes a perturbative plane gravitational wave in general relativity in 1916. After a centenary exploration, LIGO and VIRGO probe the first signal for gravitational waves from a binary black hole, which is a billion light years away from us \cite{GWs}.

 In 1925, Brinkmann found the first exact solution for gravitational wave \cite{brink}. After that, several exact solutions for gravitational waves are found \cite{exact}. Spherical wave, as a most useful model for other interactions and in linearized general relativity, is absent in general relativity for vacuum space. Several modified gravities are proposed based on theoretical and observational problems of general relativity. Usually, a modified gravity has extra parameters compared to general relativity, and it smoothly comes back to general relativity when the extra parameters vanish or reduce to some special values. In special cases, a modified gravity has non-perturbative property. These properties are especially valuable since the answer is ``quantized" for such a property, which maintains for an infinitesimal modification of general relativity. From the aspect of observations, one can distinguish general relativity from modified gravity even for one certain test invoking such a property.

  Through studies of Vaidya like spacetime in $f(R)$ gravity, we find topologically spherical vacuum wave in this modified gravity, which is strictly prohibited in general relativity. $f(R)$ gravity is a scalar tensor theory, which is conformally equal to general relativity \cite{review1}.  The topologically spherical metric, especially the flat case, is widely used in studies of AdS/CFT.

  It is easy to prove a Birkhoff-like or generalized Birkhoff theorem for topological spherical metric. For a spherical metric,
  \be
  ds^2=-b(t,r)dt^2+a(t,r)dr^2+r^2d\Omega^2,
  \label{met1}
  \en
  in which $d\Omega^2$ can be,
  \be
  d\Omega^2=dx^2+\sin^2(x)dy^2,~~~dx^2+dy^2,~~~dx^2+\sinh^2(x)dy^2,
  \en
  respectively, which denotes three different topological cases. One obtains the Ricci tensor,
  \be
  R_{01}=-\frac{1}{ra}\frac{\partial a}{\partial t}.
  \en
  For vacuum case, $R_{01}=0$ leads to $a(t,r)=a(r)$. Then one finds,
  \be
  \frac{R_{00}}{b}+\frac{R_{11}}{a}=0,
  \en
  which leads to,
  \be
  \frac{\partial b}{\partial r}\frac{1}{b}=-\frac{\partial a}{\partial r}\frac{1}{a}.
  \en
  An integration of the above equation yields
  \be
  b(t,r)=b_1(t)b_2(r).
  \en
  And $b_1(t)$ can be absorbed into the coordinate time $t$. The above calculations are independent on spatial topology. Thus we prove the Birkhoff like result for topological spherical space.
  This result shows that not only a spherical vacuum wave but also topological spherical vacuum waves are prohibited in general relativity.

  In principle, the generalized Birkhoff theorem becomes invalid in modified gravity, since a vacuum space in modified gravity is effectively equal to a space with matters in general relativity. Now we consider the
  $f(R)$ gravity.
   The action of general $f(R)$ gravity reads
   \be
   S=\frac{1}{16\pi G }\left(\int d^{4}x\sqrt{-{\rm det}(g)}~f(R)\right)+S_{m}.
  \label{action}
  \en
 Here $G$ is the Newtonian constant, $g$ is the metric tensor, and $S_m$ denotes the action of matter fields. The field equation follows the action (\ref{action}) reads,
\be
f_{R}R_{\mu \nu }-\frac{1}{2}fg_{\mu \nu }-\nabla _{\mu }\nabla
_{\nu }f_{R}+g_{\mu \nu }\square f_{R}=8\pi G T_{\mu \nu },
\label{field}
\en
 where$f_R=\frac{\partial f}{\partial R}$, and $T_{\mu \nu }$ presents the stress-energy for matter fields. For $f(R)=R^{d+1}$, the field equation becomes,
 \be
   (d+1)R^{d}R_{\mu \nu }-\frac{1}{2}R^{d+1}g_{\mu \nu }-(d+1)\nabla _{\mu }\nabla
_{\nu }R^{d}+g_{\mu \nu }(d+1)\square R^{d}=8\pi G T_{\mu\nu}.
\label{field2}
  \en
 In the following text, we set $8\pi G=1$. Only from the action form $R^{d+1}$, one may expect that all the results in this gravity come back to general relativity when $d\to 0$. However, one will see that this is not the case. In spherically symmetric static case, the solution is found by Clifton et al \cite{cb}. Then it is reconstructed by thermodynamic method in \cite{self1} and extended to higher dimensional and 3-dimensional cases \cite{self3} by thermodynamic method.  This method is further extended to obtain various solutions \cite{vari}. Some other interesting 3-dimensional solutions in extended gravity are explored in \cite{other3d}.

 Inspired by the original Vaidya metric and generalized Vaidya metric in extended gravities \cite{humass}, we write the metric ansatz as,
   \be
ds^{2}=f(u,r)du^2+2p(u,r) du dr+r^{2}\gamma _{ij}dx^{i}dx^{j},
\label{metricV1}
\en
where $\gamma _{ij}$ is the metric on a 2-dimensional
constant curvature space ${\cal N}$ with sectional curvature
  $k=\pm 1,0$, and the two-dimensional spacetime ${\cal T}$ spanned by the coordinates $(u,r)$ possesses the metric as $h_{ab}$. Comparing to the original form, this Vaidya metric has an extension that $g_{ur}$ is not a constant. One will see that this is necessary for Vaidya metric in $f(R)$ gravity.  We Assume the pure radiation as the source matter field,
  \be
  T_{ab}=\Phi (u,r) (du)_a(du)_b\doteq \Phi (u,r) (k)_a(k)_b.
  \en
  Then one obtains,
  \be
  k^a=-\frac{1}{p}\left(\frac{\partial}{\partial r}\right)^a.
  \en
  Thus the stress-energy describes a null matter propagating along $k^a$, which satisfies $k_ak^a=0$.

To obtain the field equation under the ansatz (\ref{metricV1}) is really involved but straightforward. The resulted $f(u,r)$, $p(u,r)$ and $\Phi(u,r)$ under the condition $k=1$ read,
  \be
  f(u,r)=-r^{\frac{4n-2-4n^2}{n-2}}\left(r^{\frac{2n}{n-2}}-2M(u)r^{\frac{4n^2-8n+7}{n-2}}\right),
 \label{topo1}
  \en
  \be
  p(u,r)=-\frac{1}{|2-n|}\sqrt{|7+4n-30n^2+28n^3-8n^4|}~~ r^{\frac{(n-1)(2n-1)}{2-n}},
  \label{p1}
  \en
  and,
  \be
  \Phi(u,r)=\frac{1}{3(n-2)(n-1)(4n^2-10n+7)}Pp(u,r)\frac{dM}{du},
  \en
 where,
 \be
   P=(2^{4+n}3^n-2^{5+n}3^n n+2^{3+n}3^{1+n} n^2-2^{3+n}3^n n^3+6^n n^4)\left[\frac{n^2-n}{(2n^2-2n-1)r^2}\right]^n r^{\frac{5-9n+4n^2}{n-2}}.
   \en
 In the above equations we introduce $n=d+1$, which can slightly shorten the formulae. Here $M(u)$ is an arbitrary $C^4$-function of $u$. One can check that the above solution reduces to the Vaidya solution in general relativity when $n=1$, and to the Clifton-Borrow one in outgoing Eddington-Finkelstein when $M$ is a constant. A special note is that in the reduction to the Vaidya in general relativity the factor $d$ will be cancelled in the numerator and denominator when $d\to 0$. Thus the metric ansatz (\ref{metricV1}) indeed can be treated as  generalized Eddington-Finkelstein coordinates. There is a distinct difference that $g_{ur}$ is not a constant in the pure radiation-sourced $R^{d+1}$-gravity. The property of the case $k=-1$ is similar to the case $k=1$. We just list the result here,
   \be
  f(u,r)=r^{\frac{4n-2-4n^2}{n-2}}\left(r^{\frac{2n}{n-2}}+2M(u)r^{\frac{4n^2-8n+7}{n-2}}\right),
  \label{topo2}
  \en
  \be
  p(u,r)=-\frac{1}{|2-n|}\sqrt{|7+4n-30n^2+28n^3-8n^4|}~~ r^{\frac{(n-1)(2n-1)}{2-n}},
  \en
  and,
  \be
  \Phi(u,r)=\frac{1}{3(n-2)(n-1)(4n^2-10n+7)}Pp(u,r)\frac{dM}{du}.
  \en

  Now we discuss the case $k=0$. This case leads to a non-perturbative vacuum  gravitational wave solution in $R^{d+1}$-gravity, which has no general relativity limit. The solution reads,
  \be
  f(u,r)=2M(u)r^{\frac{5-4n}{n-2}},
  \label{g00}
  \en
  \be
  p(u,r)=-\frac{1}{|2-n|}\sqrt{|7+4n-30n^2+28n^3-8n^4|}~~ r^{\frac{(n-1)(2n-1)}{2-n}},
  \label{g01}
  \en
    \be
  \Phi(u,r)=0,
  \label{phiflat}
  \en
  and the Ricci scalar,
  \be
  R=0.
  \en
  Since $T_{ab}=0$, it is a vacuum solution in $R^{d+1}$-gravity. We check that it is not the Minkowski metric in special coordinates, since
  \bea
  R_{abcd}R^{abcd}=\frac{4(n-2)^2}{(7+4n-30n^2+28n^3-8n^4)^2}r^{\frac{15-24n+8n^2}{n-2}}\left(3(47-118n+112n^2-48n^3+8n^4)r^{\frac{7}{n-2}}M(u)^2\right. \nonumber\\
  \left.+2^{1/2}(7n-2-7n^2+2n^3)r^{\frac{5n}{n-2}}p(u,r)\frac{dM}{du}\right).~~~~~~~~~~~~~~~~~
  \ena
  In previous text, we demonstrate the generalized Birkhoff theorem, which also forbids a gravitational wave in topological spherical vacuum space in general relativity. And we point out the possibility to find a such wave in modified gravity. Here we discover a typical example for gravitational wave in topological spherical vacuum space in modified gravity. Furthermore, this is a nonperturbative property of $R^{d+1}$ gravity. For what ever a tiny $d$, it exists. For an exact zero $d$, it vanishes. We will study more of this property in the following text.

  We discuss the properties of this gravitational wave in detail. First, the parameter of the graviton ray (analogy to light ray) is $u$ and thus the wave vector of this wave is,
  \be
  K_a=-(du)_a.
  \en
  To obtain the detailed properties of the null congruence of the graviton ray, we construct tetrad for this metric,
  \be
  \tau_1=du,~~\tau_2=-[g_{00}du+2g_{11}dr]\frac{1}{2},~~\tau_3=rdx,~~\tau_4=rdy.
  \en
  Under this tetrad formalism, the metric reads,
  \be
  g=-\tau_1\tau_2-\tau_2\tau_1+\tau_3^2+\tau_4^2.
  \label{tetrad1}
  \en
  The graviton congruence is geodesic congruence, since
  \be
  K^a\nabla_a K_b=0.
  \en
  Different from the case pp-wave, $K$ is not a Killing vector in view of,
  \be
  \nabla K=\frac{-1}{rg_{01}} (\tau_3^2+\tau_4^2)+\frac{-{g'}_{00}+2\dot{g}_{01}}{2g_{01}} \tau_1^2.
  \en
  Here a prime denotes derivative with respect $r$, and a dot for $u$. Since the equations become very lengthy, we directly present them in form of components of the metric in (\ref{metricV1}) and (\ref{g00},~\ref{g01}). To study the motion of the congruence, we explore its expansion, shear, and twist on sectional 2-surface $\tau_3-\tau_4$. We use a hat to denotes the projection of a tensor on the section $\tau_3-\tau_4$. The expansion, shear, and twist read,
  \be
  \hat{\theta}=\hat{g}_{ab}(\nabla^a K^b){\hat{}}=-\frac{2}{rg_{01}},
  \en
  \be
  \hat{\sigma}_{ab}=(\nabla_{(a} K_{b)}){\hat{}}-\frac{1}{2}\hat{\theta}\hat{g}_{ab}=0,
  \en
  and,
  \be
  \hat{\omega}_{ab}=(\nabla_{[a} K{_b]}){\hat{}}=0,
  \en
  respectively. Thus the graviton ray congruence is shear-free, twist-free, but expanding. This result agrees with our physical intuition for a radial gravitational wave. About the ratio of variation of $\hat{\theta}$, the Raychaudhuri equation for affine parameterized null congruence tells,
  \be
  K^a\nabla_a \hat{\theta}=-\frac{1}{2}\hat{\theta}^2-\hat{\sigma}_{ab}\hat{\sigma}^{ab}+ \hat{\omega_{ab}}\hat{\omega^{ab}}-R_{ab}K^aK^b=-\frac{2}{r^2g^2_{01}}-\frac{2g'_{01}}{rg^3_{01}}.
  \en
  Note that here the congruence has affine parameterized since $K^a\nabla_a K_b=0$. For a non affine parameterized congruence, the Raychaudhuri equation will be different.  The Raychaudhuri equations for $\hat{\sigma}_{ab}$ and $\hat{\omega}_{ab}$ are not presented here, since they are identically equal to zero all over the congruence.

  To clearly study the expansion of the graviton ray congruence, we write $\theta$ and $K^a\nabla_a \hat{\theta}$ explicitly in coordinates $u$ and $r$,
  \be
  \hat{\theta}=\frac{2|d-1|}{\sqrt{1-4d+6d^2-4d^3-8d^4}}~r^{\frac{1+2d^2}{d-1}},
  \en
  and,
  \be
  K^a\nabla_a \hat{\theta}=-\frac{2(1-d+2d^2-2d^3)}{1-4d+6d^2-4d^3-8d^4}~r^{\frac{2(1+2d^2)}{d-1}}.
  \en
  It is clear that both expansion and its ratio of variation is independent on $u$. Based on the studies of Vaidya and Kinnersley metrics, one finds that $u$ can be regarded as time in some sense. Thus we call the wave in  (\ref{g00},~\ref{g01}) steady (not standing) gravitational wave. When $d\to 0$, $\hat{\theta}$ and $K^a\nabla_a \hat{\theta}$ have well-posed limits as follows,
  \be
  \lim_{d\to 0}\hat{\theta}=\frac{2}{r},
  \en
 and,
  \be
  \lim_{d\to 0}K^a\nabla_a \hat{\theta}=-\frac{2}{r^2}.
  \en
  However, we know that $d=0$ implies general relativity, in which such topological spherical wave is forbidden. There is no wave in such case in general relativity, therefore $\hat{\theta}$ and $K^a\nabla_a \hat{\theta}$ in such space must be zero. $d=0$ is a discontinuous point for $\hat{\theta}$ and $K^a\nabla_a \hat{\theta}$. Again we see that the topological spherical wave is a non perturbative phenomenon in $R^{d+1}$ gravity.

  Then we discuss the energy transfer in this  topological spherical wave space. The energy of gravitational field is very intricate. Because of equivalence principle, stress-energy in tensor form does not exist. However, under several situations, we have to consider gravitational energy inclosed in a finite space, for example gravitational waves carry energy from remote black hole to the Earth, energy of the baryons is transferred to the gravity field in collapsing process of a star. Based on these considerations and several others, people develop the theory of quausilocal energy. Based on the inherent symmetry of the spacetime, one develops the generalized Misner-Sharp energy for $f(R)$ gravity \cite{self5}. The generalization is based on the idea related to conserved charge of conserved current. For spherical space, Kodama vector is defined as,
  \be
  O^a=-\epsilon ^{ab}\nabla_b {r},
  \en
  where $\epsilon ^{ab}$ is the Levi-Civita tensor of the 2-dimensional $u-r$ space of  (\ref{metricV1}), and $r$ is the areal coordinate. It is easy to confirm that,
  \be
  \nabla^a O^b+\nabla^b O^a=0,
  \en
  which is in analogy to the property of a time-like Killing vector in stationary space.
  We thus define a conserved charge,
  \be
  M_{MS}=-\int *(T^{ab}O_b),
  \en
  where $T_{ab}$ is the stress-energy of the spacetime and a star denotes Hodge dual operator. After some calculations, one arrives at a little involved result \cite{self5}. Invoking this result, one finds,
  \be
  2M_{MS}=\left(2M(u)\right)^{\frac{1-d}{1+2d+4d^2}},
  \en
  which is independent on $r$. The result is the same for all the 3 topologies for (\ref{topo1}), (\ref{topo2}), and (\ref{g00}). When $d=0$, it naturally comes back to Vaidya mass function in general relativity.  In the cases of $k=\pm 1$, the variation of $M(u)$ can be ascribed to the energy carried away by pure radiations, which is similar to the original Vaidya solution. However, the case $k=0$ is a pure empty space in which $T_{ab}=0$. Under this situation, one has to attribute the variation of $M(u)$ to gravitational radiations via radial gravitational waves. This case has no general relativistic limit. Again, one sees that it is a non perturbative effect.

   In addition to the argument from the energy radiation, we make further discussion about the problem ``why the metric (\ref{metricV1}) and (\ref{g00},~\ref{g01}) describes a gravitational wave". Let's review the first known exact gravitational wave solution, i.e., the Brinkmann solution,
  \be
  ds^2=\eta_{ab}+Pd\xi d\xi.
  \en
 Here $\eta_{ab}$ is the Minkowski metric in Cartesian coordinates ($t,~x,~y,~z$), $\xi=t-z$, and $P$ is a function of $\xi,~x,~y$ which satisfies,
 \be
 \frac{\partial^2 P}{\partial x^2}+ \frac{\partial^2 P}{\partial y^2}=0.
 \en
 This requirement comes from Einstein field equation. $P$ has several different forms. For example, one sets $P=f(\xi)(x^2-y^2)$, which describes a plane wave travelling along $z$-direction on a Minkowski background.

 This approach provides a clue to find our solution for gravitational wave in a different way. Besides Minkowski space, the $R^{d+1}$ gravity permits a second ground state,
  \be
ds^{2}=2pdu dr+r^{2}(dx^{2}+dy^{2}),
\label{metricG2}
\en
 where $p$ takes the form in (\ref{p1}). This metric is not Minkowski since,
 \be
 R_{ab}=-\frac{2d(1+2d)}{(1-d)r^2}(dr)_a(dr)_b.
 \en
 One confirms that (\ref{metricG2}) satisfies the field equation (\ref{field}) without source. Similar to the case of Brinkmann, one introduces a gravitational wave term propagating along areal coordinate $r$ on the background (\ref{metricG2}),
  \be
ds^{2}=q(u,r)\beta (x,y)du^2+2p du dr+r^{2}(dx^{2}+dy^{2}).
\label{metricG3}
\en
  The wave solution (\ref{g00}), (\ref{g01}), and (\ref{phiflat}), which is happen found through investigations Vaidya-like solutions, only has a longitude polarization. However, generally a gravitational wave has transverse polarizations in any gravitational theory which generalizes
  general relativity. Thus we naturally introduce the transverse polarization term $\beta (x,y)$ to find a more general wave than that of (\ref{g00}), (\ref{g01}), and (\ref{phiflat}).

 Substituting (\ref{metricG3}) into the field equation (\ref{field}) with $T_{ab}=0$, and after some involved but routine calculations, one arrives at,
 \be
 q=C(u)r^{-\frac{1-4d}{1-d}},
 \en
 and $\beta (x,y)$ is an arbitrary C-2 function of $x,~y$.
 Setting $C(u)=2M(u)$, and calling to mind $n=d+1$, one checks that the about $q$ is exactly the same as $f$ in (\ref{g00}).  Similar to the Brinkmann solution describing a wave on a Minkowski background, the metric (\ref{metricG3}) describes a gravitational wave on a background (\ref{metricG2}).

 Finally, we explore the algebraic property of this spherical wave spacetime. We write the tetrad (\ref{tetrad1}) in Newman-Penrose formalism,
 \be
 g=-\tau_1\tau_2-\tau_2\tau_1+m\bar{m}+\bar{m}m,
  \label{tetrad2}
  \en
 where,
 \be
 m=\frac{1}{\sqrt{2}}(\tau_3+i\tau_4),~\bar{m}=\frac{1}{\sqrt{2}}(\tau_3-i\tau_4).
 \en
 We find,

 \be
 \Psi_2=-C_{1342}=\frac{M(u)\beta (x,y)(1-d)(2+d)}{2(1-4d+6d^2-4d^3-8d^4)}~r^{\frac{4+2d^2}{d-1}},
 \en

 \be
 \Psi_3=-C_{1242}=-\frac{M(u)\beta (x,y)|1-d|(2+d)}{2\sqrt{2}\sqrt{1-4d+6d^2-4d^3-8d^4}}~r^{\frac{4-d}{d-1}}\left(i\frac{\partial \beta}{\partial y}+\frac{\partial \beta}{\partial x}\right),
 \en
 and,
 \be
 \Psi_1=\Psi_4=0.
 \en
 Thus it is a metric of Petrov-II. If the wave has no transverse polarizations, it reduces to a Petrov-D metric since $\Psi_3$ vanishes under this condition.


 At last, we say something about the status of $R^{d+1}$-gravity in modified gravities. $R^{d+1}$-gravity is the most pure and tractable in f(R) gravities. Even such a pure construction may explain some intricate phenomenon \cite{harko}. Furthermore, some important example in  f(R) gravity models\cite{modi} reduces to  $R^{d+1}$-gravity. For example, the Starobinsky model,
  \be
  L=R+\alpha R^2,
  \en
  reduces to $R^2$-gravity at high energy scale. In the very early universe, for example the inflation stage, it is just $R^2$-gravity. Also,  $R^{d+1}$-gravity has an interesting non-trivial black hole solution \cite{cb}.

  We concisely summarize the main results in this letter. Some phenomena of gravitational waves in f(R)-gravity are studied in \cite{gwfr}.  We find a topologically spherical gravitational wave in vacuum space in $R^{d+1}$ gravity. This type of wave is forbidden in general relativity. This is a non perturbative effect. For an infinitesimal positive or negative $d$, this wave exists. But it suddenly disappears when $d=0$. That is to say, any infinitesimal deviation from general relativity leads to such a wave. We investigate properties of the congruence of graviton rays of this wave, and find it is twist-free, shear-free, but expanding. We find this wave solution through investigations of a Vaidya-like solution in $f(R)$ gravity. We also show how to obtain it through an approach similar to the Brinkmann plane wave solution. Algebraically, it is a Petrov-II metric. It is found $d< 10^{-19}$  in the weak field approximation by using the tests in solar system. Due to our studies, even such a tiny differentiation from Einstein gravity leads to an inevitable longitude polarization for gravitational waves. At present stage, we do not have information of polarization of gravitational waves. We hope that the third generation GW detectors like the Einstein Telescope \cite{et} and Cosmic Explorer \cite{ce} find some indications for polarizations of gravitational waves.

 {\bf Acknowledgments.}
   H.Z. thanks Kai Lin, Tao Zhu, and Qingguo Huang for helpful discussions. This work is supported in part by Shandong Province Natural Science Foundation under grant No.  ZR201709220395, and the National Key Research and Development Program of China (No. 2020YFC2201400).

\end{document}